# A universal model for languages and cities, and their lifetimes

Çağlar Tuncay
Department of Physics, Middle East Technical University
06531 Ankara, Turkey
caglart@metu.edu.tr

**Abstract:** Present human languages display slightly asymmetric log-normal (Gauss) distribution for size [1-3], whereas present cities follow power law (Pareto-Zipf law)[4]. Our model considers the competition between languages and that between cities in terms of growing (multiplicative noise process)[5] and fragmentation [6]; where, relevant parameters are (naturally) different for languages and cities. We consider lifetime distribution for old and living languages and that for old and living cities. We study also the effect of random elimination (punctuation) within time evolution of languages and cities. Finally, we assume decreasing exponential distribution for cities over size with independent random amplitude and random (negative) exponent; and show that, this gives the Pareto-Zipf law.

1 **Introduction:** Our essential aim is to show that similar processes may be governing time evolution of competition between languages or cities, yet with different parameters, where the most crucial ones are two: Rate for the growth within a multiplicative noise process [5] and a factor for fragmentation [6]. It is similar to real life; origination (birth), spread and extinction takes place in terms of fragmentation and growth is due to multiplicative noise process.

Empirical criteria for our results are: i) Number of living languages and that of living cities (towns, villages, etc.) may be different; but, total size for the present time must be the same for both cases, where the mentioned size is world population. ii) World population increases exponentially (super exponentially, for recent times) with time.[3, 7] iii) At present, the biggest language (Mandarin Chinese) is spoken by about 1.025 billion people and world population (as, the prediction made by United Nations) is 6.5 billion in 2005, (and to be about 10 billion in 2050) [7]; so the ratio of the size for the biggest language to (the total size, i.e.,) world population must be (about) 1:6.5. iv) Size distribution for the present time must be slightly asymmetric log-normal for languages [1], and this must be power law -1 for the present cities (Pareto-Zipf law)[4]. It is known that, many (old) languages and (old) cities are not living at present. So, we consider random elimination in both cases (punctuation); which might have occurred due to some historical reasons, such as mass immigration, being colonized, lose of war, and wide spread illness, etc. Secondly, we aim at showing that Pareto-Zipf law may be emerging from decreasing exponential distribution of number of cities over size, where independent random numbers are used; one for amplitude and one for (negative) exponent. Following section is the model, and next one is applications and results. Last section is devoted for discussion and conclusion. We will display some more properties of the mentioned exponential functions within Appendix.

**2 Model:** For languages or cities ($i$), we have various number of ancestors ($M(0)$), all of which have random initial size $P_i(0)$. They grow in time ($t$), with a random rate ($R_i \le R$, where $R$ is universal) within a random multiplicative noise process,

$$P_i(t) = (1 + R_i)P_i(t - 1) \ , \qquad (1)$$

and if a random number is smaller than the splitting probability $H$, then the language or city ($i$) splits, with the splitting ratio (fragmentation, mutation factor) $S$, which has the following meaning: If the current number of speakers of a language or population of a city ($i$) is $P_i(t)$, $SP_i(t)$ many members form another population and $(1-S)Pi(t)$ many survive during the fragmentation at $t$. Number of languages or cities ($M(t)$) increases by one if one language or city ($i$) splits; if any two of them split at $t$, then $M(t)$ increases by two, etc. Furthermore, we assume that, if for $H \ll 1$ this random number is larger than some $G$ near 1, the language or the city becomes extinct. It is obvious that, results do not change if $1 - S$ is substituted for $S$, i.e., if the mutated and surviving members are interchanged. So, the greatest value for $S$ is 0.5, effectively. Please note that, fragmentation causes new agents to emerge (birth), and at the same time it drives them to extinction in terms of splitting, and any language or city with a member less than unity is considered as extinct.

Time evolution of (total number of human speaking all of the current languages or living in all of the current cities, i.e.,) world population is

$$W(t) = \sum^{M(t)}_{i=1} P_i(t) \ , \qquad (2)$$

where a city may actually also be a village of single house, except if we demand a minimum population.

It is obvious that, if one changes $M(0)$ to $M'(0)$ without changing origin and unit for time scale, then $M(t)$ must be changed to $M'(t)$, where $M'(t)= M(t)(M'(0)/M(0))$. Secondly, if time steps of the evolution are long enough (and the total number of time steps is small), we may have more fragmentation per unit time. If on the other hand, time steps are considerably short (for big number of total time steps), intermittency may become crucial. Moreover, there may be various other reasons for waiting periods of time in fragmentation, as well as in population growth. We assumed an intermittency factor ($H$) for splitting, which is taken same for all of languages or of cities. Please note that, $H=1=G$ gives gradual evolution, where we have a regular fragmentation with $H$ (and, with some $S$) at each time step $t$. This case is kept out of the present scope, because we consider it as (historically) unrealistic. Thirdly, we know that not all of the ancestor and old languages or cities survive, and we have many ancient (and older) languages which are not spoken any more and many of the old cities are archeological sites now. The relevant number of people might have immigrated and changed the language and city, they might have changed their language after being colonized within the same cities and they might have became totally extinct after a lost war, or after a wide spread and severe illness, etc. We count such cases as random eliminations by the rate $G$, at each $t$. We keep the (historical) reasons out of the present scope, and consider only the results; and once the reasons are kept away, the results of them may naturally be taken as random. In summary; if a random number (which is defined for each $t$) is smaller than $H$, for $H<1$, then the language or city splits (by $S$); and, if the random number is larger than some $G$ near 1, then the language or city becomes extinct; and, in between it grows (for each time step by $R$, at most). Number of languages or cities increases, decreases, or fluctuates about $M(0)$ for (relatively) big numbers for $H$ (high fragmentation) and $G$ (low elimination), for small numbers for $H$ (low fragmentation) and $G$ (high elimination), and for $H+G=1$ (equal fragmentation and elimination), respectively; out of which we regard only the first case, where we have (for $1<H+G$) increase in the number of agents and we disregard the others. Because, it is known (guessed) that the number of cities (languages) did not decrease ever in past. We try several number of ancestors $M(0)$, with sizes $P_i(0)$, for the selected case, where we assign new random growth rates for new languages or cities, which are not changed later, as well as the growth rates for ancestors are kept as same through the time evolution.

For lifetime, we simply subtract the number of time step at which a language or city is generated, from that one at which the given agent became extinct, where the agent may become extinct since its size becomes less then unity in terms of fragmentation or since it is randomly eliminated. In any case, we have two different definitions for lifetime; one is for the extinct agents and the other is for living agents (age). We may define $\mu(\tau,t)$ as the probability (density) function, where $t$ stands for the number of time steps, and $\tau$ stands for ages, and $\mu$ counts the languages or cities with age $\tau$ at $t$ (with $0<\tau\leq t$). It is obvious that, the integral of $\mu$ over $\tau$ (with $0<\tau\leq t$) gives the number of living (not all, but) generated languages or cities at $t$, which is equal to the number of total living agents ($M(t)$) minus the number of living ancestors, which goes zero with $t\rightarrow\infty$.

Please note that, the introduced parameters have units involving time, and our time unit is arbitrary. After some period of evolution in time we (reaching the present) stop the simulation and calculate probability distribution function (PDF) for size, and for some other functions such as extinction frequency, lifetime, etc., (for languages or cities). Number of interaction tours may be chosen as arbitrary (without following historical time, since we do not have historical data), with different time units; and, the parameters (with units) may be refined accordingly. Yet, in most cases relative values (with respect to other cases; population growth rate in different runs, for example) and ratios of the parameters (the ratio of $S$ to $H$, for example) are important.

**3 Applications and results:** Languages and cities are man made systems; and, language is not a physical quantity, whereas city is a physical quantity. Yet, we find many similarities between them. We first consider languages (sect. 3.1), later we study cities (sect. 3.2), in section 3.3 we compare the results for languages and cities; and finally, in section 3.4, we utilize exponential (decreasing) functions with two independent random numbers; one for the amplitude and one for the (negative) exponent, where our essential aim is to show that the mentioned sum gives Pareto-Zipf power law for size distribution.

*3.1 Languages:* In our initial world (at our t=0) we have $M(0)$ (=100) many languages, each of which is spoken by randomly chosen $P_i(0)$ ($\leq$10,000) people; thus, the initial world population ($W(0)$) is about 500,000, since average of the (homogeneous) random numbers between zero and unity is 0.5. Thus we assume power law zero for the initial distribution of languages over size, i.e., $M(P)\propto P^0$, for initial parameters $M(0)$ and $P_i(0)$.

It is obvious that, we may not set our time origin correctly, because not a real data for the initial time is available to match with. Yet, we may assume a homogeneous distribution of size over initial languages. In fact, we tried many smooth (Gauss, exponential, etc.) initial distributions (not shown); and, all of them undergone similar time evolutions within about 3000 time steps, under the present process of random multiplication for growth, and (random) fragmentation for spread and origination and extinction, where we utilized various combinations of the relevant parameters, including $H$ and $G$. We tried also delta distribution, which is equivalent to assuming single ancestor, for the initial case; it also evolved into a slightly asymmetric log-normal one (with different set of parameters, certainly; not shown) in time. In all of them, we observe that, the language distribution at present is independent of initial distributions, disregarding some extra ordinary ones.

*Evolution without elimination* ($G$=1): Number of languages ($M(t)$) and population of world ($W(t)$) increases exponentially in time, as shown in Figure 1, where the arrow for languages (dashed line) has the slope 0.000095, and that for world population (solid line) has slope 0.002530, where the parameters are: $R$=0.00752 and $H$=0.0022, with $S$=0.49999 (others are same as before).

Within the present section we utilize 3000 time steps, as shown in Fig.1, where we consider the present time (the year 2000) as $t$=1920, which is indicated by a two sided arrow in Fig.1. We show the portion for 1920 <$t$<3000, because we will make some comments (predictions) for languages about future (and also for cities; since there are close relations between them). Please note that, at $t$=1920 (present time, the year 2000) we have 7,168 languages and the world population is about 10 billion (Fig. 1).

Figure 2 is the time evolution of size distribution of languages (PDF), all of which split and grow by the same parameters, i.e., $S$, $R$, $H$ (all same as declared in the third paragraph of this section), and $G$=1 (thus, we do not have abrupt (punctuated) elimination of languages here). Lower plot is historical ($t$=270, open circles) and upper one (squares) is for the present time ($t$=1920). In Fig. 2 we observe that; initial languages spread in number by fragmentation, then we have new languages, which split from the previous bigger ones (right). New languages emerge also from the small ones (left). And all of them grow in size in terms of a random multiplicative noise process. Growing rates are similar; yet, bigger ones grow (if they do) in bigger amounts at each time step. We observe that, small languages grew from few decades to more (yet few) decades (with a time period of 1650 (=1920-270) time steps); whereas, the biggest language grew (within the same time period) from few tens of thousands to a billion as we observe at right end of Fig. 2 (Chinese, in reality). In the meantime world population grew (exponentially) from (about) half million to about 10 billion; and number of languages increased (exponentially) from (about) 197 to (about) 7,168. We have parabolic fit (dashed line) in Fig. 2, which indicates that the present distribution is slightly asymmetric log-normal. (For the empirical distribution of the present languages; see, [1, and 2-3].)

We may summary our results, in some of which we utilized several other combinations of the declared parameters (not shown) as: 1- For many combinations of parameters, we get asymmetric log-normal distributions (for example, for $H$=0.003 (with $R$=0.00752, and $S$=0.1) about 70,000 languages (not 7,000) may outcome; not shown), where asymmetry may be slight or more depending upon the parameters. On the other hand, as $M(0)$ increases without changing other parameters, top of the plot for $M(P)$ rises up; and as $P_i(0)$ increases, width and the area (i.e., world population $W(t)$) under the plot increases, all for a given number of time steps. Please note that, $M(0)$ and $W(0)$ (i.e., sum of $P_i(0)$ over $i$ for $1 \leq i \leq M(0)$) is the amplitude of the (asymptotically) exponential function $M(t)$ and $W(t)$, respectively (not shown). 2- Increase in $H$, lifts the left end and top of the plot for $M(P)$ and widens it for a given $W(t)$, since we have more generated agents (due to high fragmentation) per time step. 3- Increase in $R$, extends the plot for $M(P)$ along two directions; i.e., it widens the plot at left (small languages) and at right (big languages) end, since we have more rapid growth in every language, where mutated (split) abundance also increases. In short; we obtain various plots, which are more or less (or almost) same as the empirical data for the present languages, with different parameters. Our relevant comment is that, language distribution at present time may not be unique and similar ones may be obtained with various combinations of parameters for initial conditions and these for evolution, where range of each parameter used is considerably wide. We will revisit this issue in section 3.3. 4- One important observation that we make within our results is for $1920<t$, i.e. for future: For a given asymmetric log-normal size distribution of languages for the present time; as $t$ increases, asymmetry increases (where we keep all of the given the parameters as same, and increase $t$). Small languages end rises up with decrease in local slope with time; in the meantime local slope for big languages end becomes more and more steep. In short, small languages end becomes the top of the plot for $M(P)$, and big languages end turns to be power law with exponent about negative unity (see, Figure 2, the plot in open squares, where the slope for plot in solid squares (at tail) is –0.7). Thus, we have power law with exponent zero at the beginning, we have power law with exponent (about) negative unity for future, and intermediately i.e., at the present time we have log-normal for size distribution. We will revisit the current issue in further paragraphs of this section and in sections 3.2-3.4, and 4.

Time distribution of languages, is another important issue, besides the size distribution. We consider here lifetime for the extinct languages and ages of the present ones. Figure 3 is the lifetimes of languages at present ($t$=1920, thick solid), past ($t$<1920, thin solid) and in future ($1920<t$, dashed). (Parameters are same as before.) Here, we observe that, many ancestors survive in all $t$; only two ancestor languages are found as mutated or having become extinct before $t$=1920. Secondly, generated languages are distributed exponentially (decreasing) with age, in both cases. Inset is the number of speakers for each of the living language at present time (which evolved out of the initial ones with random size.). The biggest present language (Chinese) is spoken by little more than one billion (1G, in the inset) human.

We obtained many similar results for the same parameters; yet, for different number of time steps; where, we observed that, the (negative) exponent remained same in results. Secondly, we tried the same number of time steps (as the one utilized within the previous case) and, we change other parameters. Considering these (new) parameters, which give similar PDF for the present empirical distribution of languages over size; we observed similar exponential behavior for distribution of emerged languages over age, where the (negative) exponents depend on the parameters. Consecutively, we may state that, generated languages are distributed (decreasing) exponentially (independent of the parameters used) with age, and (negative) exponents depend on the parameters. It is obvious that, increasing $H$ (directly, and increasing $R$ indirectly; since, any size hardly becomes less than unity for big $R$) promotes many new languages to emerge. As a result, linear portions of the plots (left end in Fig. 4 which has logarithmic vertical axis for the number) becomes steeper; which means that, the exponents increase in absolute value. There, we may consider the situation as scaling the (horizontal) time axis; where, increasing $H$ and $R$, amounts to increasing the unit for a time step, and decreasing the number of time steps from the initial time ($t$=0) up to the present, and vice versa.

*Elimination* ($G$<1) plays a role, which is opposite to that of fragmentation ($H$) and growth ($R$) in evolution; where, $H$ and $R$ develop the evolution forward, and $G$ recedes. So the present competition turns out to be one between $H$ and $R$, and $G$. Number of languages are essentially determined by $H$ and $G$, where two criteria are crucial: For a given number of time steps, $R$, and $M(0)$, etc., there is a critical value for $G$; where, for $G_{critical}<G \cong 1$ languages survive, and for smaller values of $G$ (i.e., if $G \cong G_{critical}$) languages may become extinct totally. Secondly, sum of $H$ and $G$ is a decisive parameter for the evolution: If for a given $G$ (with $G_{critical}<G$), $H+G=1$, then the number of languages does not increase and does not decrease, but oscillates about $M(0)$, since (almost) same amount of languages emerges (by $H$) and becomes extinct (by $G$) at each time step, and we have intermediate elimination. On the other hand, if $H+G<1$, then languages decrease in number with time and we have high (strong, heavy) elimination. Only for $1<H+G$, (with $G\neq 1$) we have low (weak, light) elimination of languages, where the number increases (yet, slowly with respect to the case for $G$=1). It is obvious that, elimination slows down the evolution. It amounts to decreasing the unit for a time step, and increasing the number of time steps from the initial time ($t$=0) up to the present, and vice versa. So we need more time to arrive at a given configuration of languages; i.e., total number, size, total size PDF, etc.

Now we introduce $G$=0.9996, with $H$=0.0022 (as before), where $1<H+G$, and we have light elimination. We have the parameters (same, as before) for $M(0)$ (=100, number of ancestors) and $P_i(0)$ (≤10,000, number of initial speakers) which grows with the rate $R$=0.00752 (as before, Eq. (1)). At the end of 1920 time steps we have $W$ about 13 billion and $M$ about 3030 (Eq. (2)), where it may be observed that light elimination is effective on the number of languages rather than on world population. World population ($W(t)$) and number of languages ($M(t)$) increases exponentially (disregarding the abrupt and temporary recessions, not shown), with exponents 0.00235, and $7.8\times10^{-4}$, respectively. Figure 4 is the resulting size and the inset is time distribution of languages at t=1920 (solid squares, for size); where PDF is smaller in height with respect to the one (solid squares) in Fig. 2. If we increase $H$ to 0.0028 (keeping other parameters same as before) we observe a different picture (not shown); where, small size end of the PDF plot is higher, since fragmentation is higher; yet, we do not have (small) languages spoken by few people. Many small languages are now spoken by some hundred or thousand human, at minimum. Languages move towards right on the plot (PDF) for size as their speakers increase in number; and they move towards left as they fragment, and become extinct at the left end when they are spoken by no human.

Log-normal plot in open squares for t=3000 (future) in Fig. 4 is clearly asymmetric. It involves a long tail for big sizes, and a short arm for small sizes, top of the plot is for about 100,000 speakers, etc. Please note that, the plot with open squares in Fig. 2 is for $G$=1, and for $t$=3000 (for future), and the only different parameter here, with respect to the corresponding plot in Fig. 2 is $G$. Initial conditions are similar (not same, because we have random $P_i(0)$) in both cases, but evolution parameters are different; yet, the plots are similar in many aspects. First of all, in both cases we have higher left arm (for small languages) with respect to the plots for the present time; which implies that, present small languages will decay in time, where punctuation increases the rate for decay. Secondly, we have (decreasing) exponential time distribution in both cases, and punctuation ($G \neq 1$) decreases the exponent in absolute value. As a result of punctuation, we may expect more people speaking a given language in future (due to some mechanism(s) else than fragmentation; for example, change of languages in terms of physical or cultural colonization, etc.), since top of the plot for future in Fig. 4 is lower than top of the corresponding plot in Fig. 2. On the other hand, as we mentioned previously (two paragraphs before) that, presence of $G$<1 amounts to increasing the unit for time steps. So, the plot in solid squares in Fig. 4 may be viewed as one for future with smaller parameters for $H$, and $R$. More clearly; if one utilizes bigger parameters for $H$, $R$, he may obtain the target configuration for languages for smaller $t$, if, $G$=1. And, since $G$<1 eliminates $M(t)$ (randomly) we need bigger parameters for $H$ and $R$, to mimic the size distribution, world population, size of the biggest language, and other empirical data. As a result, size distribution comes out with long tail for big size, and with short arm for small size. Moreover, sizes for small languages increase; i.e., we do not have languages spoken by few human, any more.

As time evolves, top of the PDF for size moves towards left, due to fragmentation ($H$), and tail becomes more and more crowded due to increase in population ($R$). As a result, we have two configurations: For some $H$ (and $R$), we have (local) horizontal behavior for small size (saturation) and we have a decline for big size, which becomes roughly straight (in many of our runs, not shown). Secondly, for bigger $H$ (without changing $R$), top of the plot rises at small language end, and we have (roughly) straight distribution for many big languages in log-log axes, which means that we have power law (see, paragraph 6 of sect. 3.1).

We had already considered age distribution of the generated languages as (decreasing) exponentials in time for $G$=1, where age distribution for ancestor languages is delta function, i.e., same (=1920) for all of the ancestors. And now, we have similar exponential time distribution for $G$<1, yet with smaller exponent in absolute value (not shown). Secondly, for $G$<1 number of living ancestors decreases in time with high fragmentation (not shown) and with the same $H$ for fragmentation as before (not shown), where we have only 32 ancestor languages present at $t$=1920, for $G$<1, with $H$=0.0022.

Please note that, ancestor languages may fragment at any time, $t$<1920; and they may give birth to new languages, where size may vary in time for ancestor and generated languages. We count these languages as ancestor, if their lifetime is 1920 (for 1920 time steps). Figure 5 is the (time) distribution for lifetime of the languages, which became extinct; and, the inset is the relative frequency of events for extinction, all with $G$<1, where other parameters are same as before.

We observe that the distribution for lifetime for extinct languages are also (decreasing) exponential, with exponent 0.001 (as given by an exponential fit, not shown). As a result, more languages become extinct in their youth and less ones become extinct as they become old. In other words, languages become extinct either with short lifetime (soon after their generation), or they hardly become extinct later and live long (which may be considered as a kind of natural selection). It may be noted that, long living languages have many speakers, and they hardly become extinct in terms of splitting.

Certainly, we do not have any empirical record for time distribution of languages to compare our results. Yet, the number of time steps we utilized (=1920) may correspond to (roughly) last 20,000 years of history; with the initial conditions $M(0)$ (=100) and $W(0)$ (=500,000). So, our time unit may correspond to about a decade in year; and, 10 to 100 time steps (i.e., century to millennium) for a given (ancient) language to become extinct may be considered as reasonable. We count the number of extinct languages in Fig. 5 as 618 (about 10% of the living ones), which may also be considered as reasonable.

Inset in Fig. 5 shows the ratio of languages which became extinct to the ones living at any $t$ ($0<t\leq 1920$), where we observe cascades within extinctions, which decay in time. And, as time evolves, cascades disappear and we have a constant noise, for which we have the present (slightly asymmetric) log-normal size distribution ($M(P)$).

As the final remark of this section we note that, the number of time steps which we used within the runs are 3000, 2430, 1920, and 270; and, they go in squares; i.e. with $30m^2$, where m=10, 9, 8 and 3, respectively. It is because, we observed that size distribution develops in (roughly) equal amounts in log-log plot; if, we increase our maximum number of time steps in squares. For example, consider Fig. 3; when we plot PDF for t=480 ($m$=4), 750 ($m$=5), 1080 ($m$=6), and 1470 ($m$=7) they fill in space between the plot for $t$=270 and $t$=1920 almost uniformly. And, if we continue for $m$=9, and 10, the mentioned uniformity survives. We think that, evolution of language distribution for size may be running in steps, which are proportional to square of time periods. (Also; see, Figure 7, where number of ancestors decreases (roughly) uniformly with square of time.)

*3.2 Cities:* We claim that, initial world is not (much) relevant for the present size configuration of cities. Moreover, we may obtain similar configurations for various evolution parameters. In Figure 6, we display three PDF for size distributions, with different initial and evolution parameters, where the maximum population for ancestors ($P$=10,000), and splitting factor ($S$=0.1), is same (with, $G$=1) for all. Please note that, $S$ is not the maximum value (which is 0.5, effectively), since we think that any number of human may change their language, yet a city may split only in small portions. On the other hand, it is known that, minimum size for a language is unity but it is not defined for a city. A city may actually also be a village of single house, except if we demand a minimum population. For that reason we considered several minimum populations for a city. In Figure 6, slopes for linear fits (dashed lines) are about: 1) -1 (squares) 2) -1 (circles) and 3) -0.85 (stars). And, utilized parameters are: 1) (squares) $M(0)$=100, $R$=0.005, $H$ =0.01, number of time steps=2430, and minimum population for a city =626); 2) (circles), $M(0)$=100, $R$=0.0062, $H$ =0.02, number of time steps=3000, and minimum population for a city=740; and, 3) (stars) $M(0)$=10, $R$=0.005, $H$ =0.015, number of time steps=1470, and minimum population for a city =1. We observed many (other) power laws about minus unity for several (other) combinations of parameters (not shown), where we observe also that; a very wide range of parameters give the same Pareto-Zipf law. And we obtain (decreasing) exponential time distributions for lifetime of extinct cities and ages of the livings ones for big $M(0)$; and for $M(0)$ less than or equal to (about) 10, the mentioned (decreasing) exponential distributions for lifetime changes in form. Please see Figs. 3 and 4 for similar distributions for languages, and note that exponents are same for different $t$, which may be utilized to predict the terminal number for cities, and that for languages under some assumptions, as we will mention in section 3.3.

At present it is obvious that $W(t)$ and $M(t)$ increases exponentially, due to $R$ and $H$, respectively, since $G$=1. For $G_{critical}<G<1$, increase in $M(t)$ lessens, and for $G_{critical}<G$, and $G+H=1$ we have oscillations in $M(t)$ about $M(0)$, which is not a realistic case, as the one for $G_{critical}\cong G<1$, where we have few cities at the end, and we do not consider them here.

*3.3 Languages and cities:* It is worth to emphasize that, languages and cities are no longer the same and that for some intermediate times the cities obey the 1/size law, while the languages obey the log-normal distribution, for the same (or better, similar) set of parameters.

As we mentioned in the previous section, time distribution of cities (lifetime for extinct cities and ages of the livings ones) are decreasing exponentials (disregarding the cases for small number of ancestors) as shown in Figure 7. Simple probability (density) functions are also exponential (not shown), which means that cities occupy time distribution plots (Figs. 3, 4, and 7) in exponential order; more cities for small $t$, and less cities for big $t$, for a given number of time steps in all. Inset of Fig. 7 is simple probability (density) function of living cities at $t$=1920, which is decreasing exponential with the rate (exponent) of 0.00185 per time step, where the number of living ancestor cities is 37 at present as shown by a solid square, which is far from the curve; and they ($M(0)$, say about 1000 in number, at most) can be neglected within some hundred thousand cities at present, if ancestors have small sizes. On the other hand, we know that, size distribution for languages is not power law, for the present time. Yet, they display decreasing exponential time distributions (lifetime of extinct languages and ages of the livings ones; disregarding the cases for small number of ancestors) similar to the ones for cities. For random elimination case (punctuation) for languages (with $G$=0.9988, $H$=0.0022, i.e., light elimination for $R$= 0.0082) simple probability density function (not shown, similar to the one in the inset of Fig. 7) is decreasing exponential with exponent (about) 0.0017, where the number for the youngest languages (i.e. these emerged within the last 19.2 time steps, since we divide (horizontal) time axis into equal 100 parts to calculate probability) is 2430 for 1920 time steps, where decrease in the number of living ancestors in time (from hundred to about ten) is observable.

Now, we may represent the probability (density) function by $\mu(\tau,t)$, where $t$ stands for the number of time steps, and $\tau$ stands for ages of the languages at $t$, and $\mu$ counts the languages with age $\tau$ at $t$ (with $0<\tau\leq t$). It is obvious that, the integral of $\mu$ over $\tau$ (with $0<\tau\leq t$) gives the number of living (not all, but) generated languages at $t$, which is equal to the number of total living languages ($M(t)$) minus the number of living ancestors, and the number of living ancestors goes zero with t→∞.

In Fig. 7 (and in many others, not shown) we observe that exponent ($\alpha$, say) of $\mu(\tau,t)$ is independent of $t$, for languages and cities. Assuming that, the exponent ($\alpha$=0.00185) for cities, will remain as constant for t→∞, we may predict M(t→∞): A simple integration of $\mu(\tau,t)=\mu(1,t)\exp(-\alpha\tau)$ over $\tau$ (with $0<\tau\leq t$) can be performed for $M(t)$, since living ancestors decay (in terms of fragmentation and random elimination) with time ($M(t)=(\mu(1,t)/\alpha)\exp(-\alpha t)$). The crucial parameter for the result is $\mu(1,t)$, i.e. number of newly generated languages or newly established cities at $\tau$=1, for each $t$. It is obvious that $\mu(1,t)$ can not be constant or a decreasing function with $t$. Because we have elimination with $G$<1, and number of eliminated languages or cities increases with $t$; languages change and some of them become useless, and cities become old and ruined in time. $\mu(1,t)$ may be an increasing function in time, due to non zero fragmentation ($H$). If $\mu(1,t)$ increases linearly with time we get saturation in M(t→∞), which is not contradiction, since we have elimination. If $\mu(1,t)$ increases exponentially with the exponent $\beta$, we have saturation in M(t→∞) for $\beta=\alpha$; and we have M(t→∞)=0 for $\beta<\alpha$. Since, we know that, at present $M(t)$ increases with time, we may guess that $\alpha<\beta$ ; i.e., we invent new languages (and establish new cities), which increase either linearly or exponentially (with the exponent bigger than 0.0017 for languages and bigger than 0.00185 for cities) in time. One may guess that M(t→∞) will saturate in reality, since human population (for languages) and surface area of earth (for languages) is limited.

On the other hand, it is known that, cities were states in past; yet, it is not certain whether one language was spoken per city (state), or vice versa. For the present, we have about 7000 languages at present for a world population which is little less than 10 billion (as mentioned within the first paragraph of sect. 3.1). For the size distribution designated by number 1 (squares) in Fig. 6, we have about 110 thousand cities with a world population which is about 5 billion (not shown). So, one may say that, a language is spoken within 15 city on the average, at present. Yet, it is not meaningful because of several reasons; one of which is that, we counted these as city, which have 626 citizens at minimum. And we do not know totally how many human are living in towns and villages, etc. But, it is obvious that we have much more cities than the languages in the present world. We have many metropolitan cities in the present world, where only one language is spoken by big majorities. And, we have many cities per country, where the number of languages is smaller than the number of cities, etc. It is certain that, ratio of the number of total languages to that of cities is less than unity, at present.

Thus far, we considered within various paragraphs that, languages and cities (as products of the same humans) are very similar to each other in many respects. Yet, languages and cities are no longer the same; and for some intermediate times the cities obey the 1/size law, while the languages obey the log-normal distribution, for the same (or better, similar) set of parameters. We predict that, log-normal size distribution may turn into power law with the same, parameters if (only) the number of time steps is increased (i.e., in future). Moreover, languages and cities have (decreasing) exponential time distributions (lifetime for extinct cities and age for the livings ones).

In the next section, we will consider power law distribution for cities over size, where we will utilize big number sum of (decreasing) exponential functions with independent random numbers for amplitude and exponent, and show that, the sum gives power law -1.

*3.4 Sum of exponential functions with random and independent amplitude and exponent*: Let's consider exponential functions ($z_{jk}(x)$), where the amplitude and exponent are random and independent; for a real or complex number $b$,

$$z_{jk}(x) = A_j \exp(B_k b x) \quad . \qquad (3)$$

In Eq. (3), $x$ is any independent variable; $A_j$ is one random number $A_{min} \leq A_j \leq A_{max}$ and $B_k$ is another random number $B_{min} \leq B_k \leq B_{max}$ which are independent of each other. In other words, we select $A$ and $B$, from different sets of random numbers, between which there is no connection. It is obvious that, real $b$ may be taken as a scaling factor for $x$, or for $B_k$; i.e., one may select $b$=1 for a unit of $x$-axis, and one may select $b \neq 1$ for a scaled unit of $x$-axis with $x \rightarrow bx$. And similarly; one may select $b$=1 for some $B_{min}$ and $B_{max}$, and one may select $b \neq 1$ for $B_{min} \rightarrow bB_{min}$ and $B_{max} \rightarrow bB_{max}$. Now we take $b$ in Eq. (3) as negative for positive $B_k$, ($b \rightarrow -b$), thus we will have negative exponents.

Within a system we may have several physical quantities, which vary in proportion to $z_{jk}(x)$, or to derivatives and integrals of $z_{jk}(x)$. Let's consider one such function $y_{jk}$(x), which is proportional to $z_{jk}(x)$, i.e., $y_{jk}$(x)= $z_{jk}(x)$, where the constant for proportion is taken as unity, for simplicity. We define,

$$Y(x) = \sum_j^J \sum_k^K y_{jk}(x)/JK \quad , \qquad (4)$$

with some big yet finite $J$ and $K$, where the order of sum is not important due to independence of the random numbers, and random numbers are generated homogeneously.

Inset of Figure 8 is for $y_{jk}(x)$ (Eq. (3)) with b=0.1 and $A_{min}$= $B_{min}$ =0.0, $A_{max}$ = $B_{max}$=1.0, which may be considered as size distribution for languages or cities along $x$-axis, where $x$ counts the agents (i.e., we place an agent ($i$+1) at $x$+1, if it is generated right after an agent ($i$) which is placed at $x$). Please note the present ancestors for small $x$. For the sum in Eq. (4), one may select the random numbers same for all $x$, and one may select new (different) random numbers at each $x$, where all of the random numbers are completely independent ($A_i$ is one random number, $A_{min} \leq A_i \leq A_{max}$; and $B_i$ is another random number, $B_{min} \leq B_j \leq B_{max}$) in all. Figure 8 displays $Y(x)$ (Eq. (4)) with $J=K=1000$, for $A_{min}= B_{min} =0.0$, $A_{max} \leq 1.0$, $B_{max} \leq 1.0$, where we have no fluctuation in $Y$ for the random numbers selected same for each $x$ (dashed thick line) and fluctuation in $Y$ increases in magnitude, as $x$ increases (fluctuating plot) for the random numbers (re-)selected for each $x$. And, in both cases we have power law; as the arrow with slope -1 indicates in log-log axes.

Obviously, the important issue in Fig. 8 is that $Y(x)$ is a power function for large $x$ with the power of minus unity, i.e.,

$$Y(x)=\sum_{\mathrm{j}}^{\mathrm{J}}\sum_{\mathrm{k}}^{\mathrm{K}} y_{jk}(x)/JK \propto x^{-1} \quad , \qquad (5)$$

which is similar to Pareto-Zipf law, if $x$ stands for city population. We may state that, if the probability of finding a city with population about $x$ (say, $x \pm dx$) decreases exponentially with independent random amplitude and exponent, then the sum of mentioned probabilities gives the famous power law. In other words, reason for the mentioned power law is the randomness in amplitude and (negative) exponent of the probability of finding a city with population about $x$ (say, $x \pm \mathrm{d}x$). Clearly, Eq. (5) is valid for every situation (competition between cities, that between languages, and similarly between speciation and extinction, etc.), if it satisfies the given conditions (i.e., random and positive amplitudes, independently random and negative exponents).

For homogeneously (yet, independent) distributions for the random numbers $A_j$ (for the amplitudes) and $B_k$ (for the exponents), with $A_{min} \leq A_j < A_{max}$ and $B_{min} \leq B_k \leq B_{max}$, one may utilize the general theorem of central limit with large $N$ or just convert the double sum into a double integral over $A$ ($A_{\mathrm{j}} \rightarrow A$) and $B$ ($B_{\mathrm{k}} \rightarrow B$), which vary linearly (since the random numbers are homogeneously distributed) between the extrema to obtain the following equalities:

$$\begin{aligned} Y(x)&=\sum_{\mathrm{j}}^{\mathrm{J}}\sum_{\mathrm{k}}^{\mathrm{K}} y_{jk}(x)/JK \\ &\propto -(A^2_{max} - A^2_{min})[\exp(-B_{max}bx) - \exp(-B_{min}bx)]/(2bx) \quad , \qquad (6) \end{aligned}$$

and, for $B_{min} =0$, we have

$$Y(x) \propto -(A^2_{max} - A^2_{min})[\exp(-B_{max}bx) - 1]/(2bx) \quad , \ (7)$$

which may be further simplified, for $1 \ll 2bx$;

$$Y(x) \propto [(A^2_{max} - A^2_{min}) /2b]\, x^{-1} \quad , \qquad (8)$$

since, the exponential term approaches to zero as $x$ increases. And the result is power law, with power minus one (Eq. 5). It is obvious that, $B_{max}$ may always be taken as unity, and $b$ may be varied accordingly. We may state that, our analysis (Eqs. (5)-(8)) is general for any set of independent random numbers for amplitudes (provided $A_{max}$ and $A_{min}$ is finite) and that for exponents (provided $B_{min}= 0$, where $B_{max} =1$ may be taken after changing $b$ accordingly). (For some other applications of the present issue (in biology, for example); see, [8].)

As we mentioned at the beginning of this section that, $b$ in Eqs. (3)-(8) may be real or complex. It is obvious that, for a complex (specifically, purely imaginary) $b$, $Y(x)$ may be expressed in terms of sinus and cosinus functions, so we get oscillatory behavior in $M(t)$ (Eq. (2)) for $G+H=1$ for example (and, similarly, in time series for price in economy, etc.). Furthermore, $y_{jk}(x)$ in Eq. (4) may be proportional to first (second, etc) derivative of $z_{jk}(x)$ in Eq. (3), or to integral of it; where, we have more interesting cases, since then we have more than two random numbers in $y_{jk}(x)$. We consider these cases within Appendix.

**Appendix**

*Power law for harmonic oscillation:* For complex $b$ (i.e., $b \rightarrow \pm ib$), $y_{jk}(x) = A_j \exp(\pm B_k bx)$, which may be re-written as a linear combination of sine and cosine; $y_{jk}(x) = A_j \sin(B_k bx) + A'_j \cos(B_k bx)$. Then we have

$$Y(x)=\sum_j^J \sum_k^K [A_j \sin(B_k bx) + A'_j \cos(B_k bx)]/JK \quad , \qquad (9)$$

which is similar to Eq. (9) in [9]. It is obvious that, not $Y(x)$ in Eq. (9) but its envelope has power law -1, as the following equation implies

$$Y(x) \propto \pm(A^2_{max} - A^2_{min})[\sin(B_{max}bx) \pm \cos(B_{min}bx)]/(2bx) \quad . \qquad (10)$$

*Power law with higher power:* Now, let's suppose that $y_{jk}(x)$ is proportional to $d(z_{jk}(x))/dx$, then we have three (at a first glance, it may be seen as two) random numbers (two of which are independent) in $y_{jk}(x) = bA_jB_k\exp(B_kbx)$, after taking the proportionality constant as unity. One may show (after a similar analysis, Eqs. (3)-(8)) that, $Y(x) \propto x^{-2}$. Similarly, if $y_{jk}(x)$ is proportional to $d^2(z_{jk}(x))/dx^2$, then we have $Y(x) \propto x^{-3}$, etc. (For similar distribution functions for price in economy; where, $x$ stands for the relative price change (more precisely for the change in the logarithm of the price) one may see [10], and relevant references therein.)

**4 Discussion and conclusion:** Starting with random initial conditions and utilizing many random parameters for evolution, we obtained many regularities (for size and time distributions, etc.) under the process of random multiplicative noise for growth and random fragmentation for generation and extinction of languages or cities, with no further random elimination ($G$=1). We find that, results are independent of initial conditions, disregarding some extra ordinary ones. If we have $G \neq 1$, we need longer time to mimic a target configuration, where new generated languages or cities may be inserted, in terms of fragmentation, provided $G_{critical} \leq G$, and $1 \leq G+H$. Furthermore, many languages or cities become extinct in their youth and less ones become extinct as they become old. In other words, languages or cities become extinct either with short lifetime (soon after their generation), or they hardly become extinct later and live long (which may be considered as a kind of natural selection).

It is known that, logistic maps (Eq. (1)) are dynamic, and under some circumstances they may become chaotic. We predict that, time distribution for languages or cities are (decreasing) exponential and log-normal size distribution for languages at present may turn into power law in future. Power law -1 for size distribution of present cities, may be result of (decreasing) exponential distribution of their probability over size. It is worth to mention that, finite sum of (decreasing) exponential functions with two independent random numbers (one for the exponent, and one for the amplitude) may be useful to express many relevant phenomena, as well in biology and economy.

In summary, languages and cities are not the same and for some intermediate times the cities obey the 1/size law, while the languages obey the log-normal distribution; where the decisive parameters are fragmentation rate ($H$) and splitting factor ($S$): We have smaller ($H$) and bigger ($S$) for languages with respect to these for cities, for the same population growth rate ($R$, which governs the world population, which is the same for languages and cities) and for the same set of other parameters.

**FIGURES**

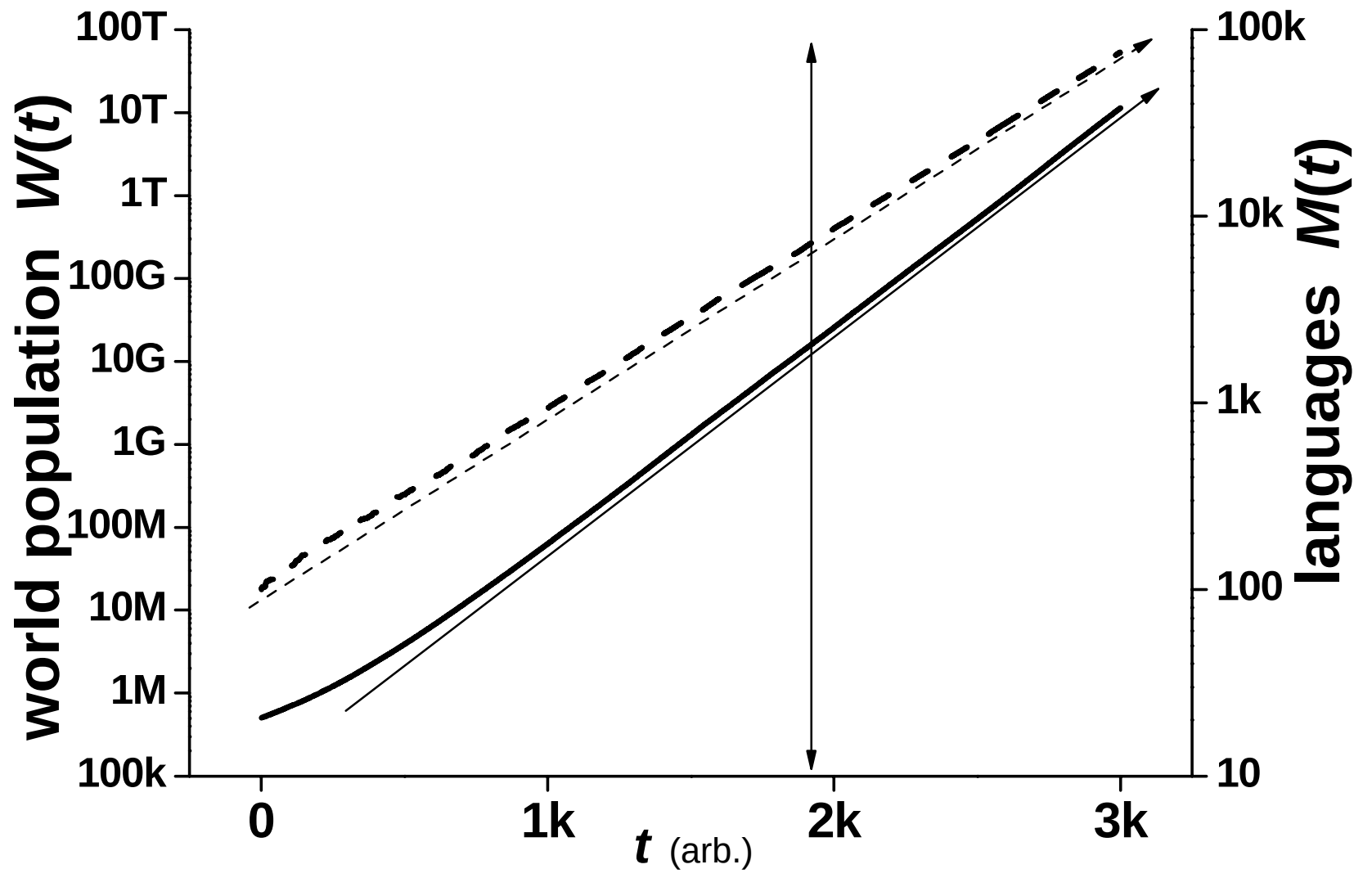

**Figure 1** Time evolution of world population ($W(t)$) (solid) and that of the number of languages ($M(t)$), Eqn. (1)) (dashed). Vertical (two sided) arrow (at t=1920) indicates the present time (the year 2000), and the slope of the arrow for the world population (left axis) and that for the number of languages (right axis) is 0.002530 and 0.000095, respectively. Please note that the vertical axes are logarithmic.

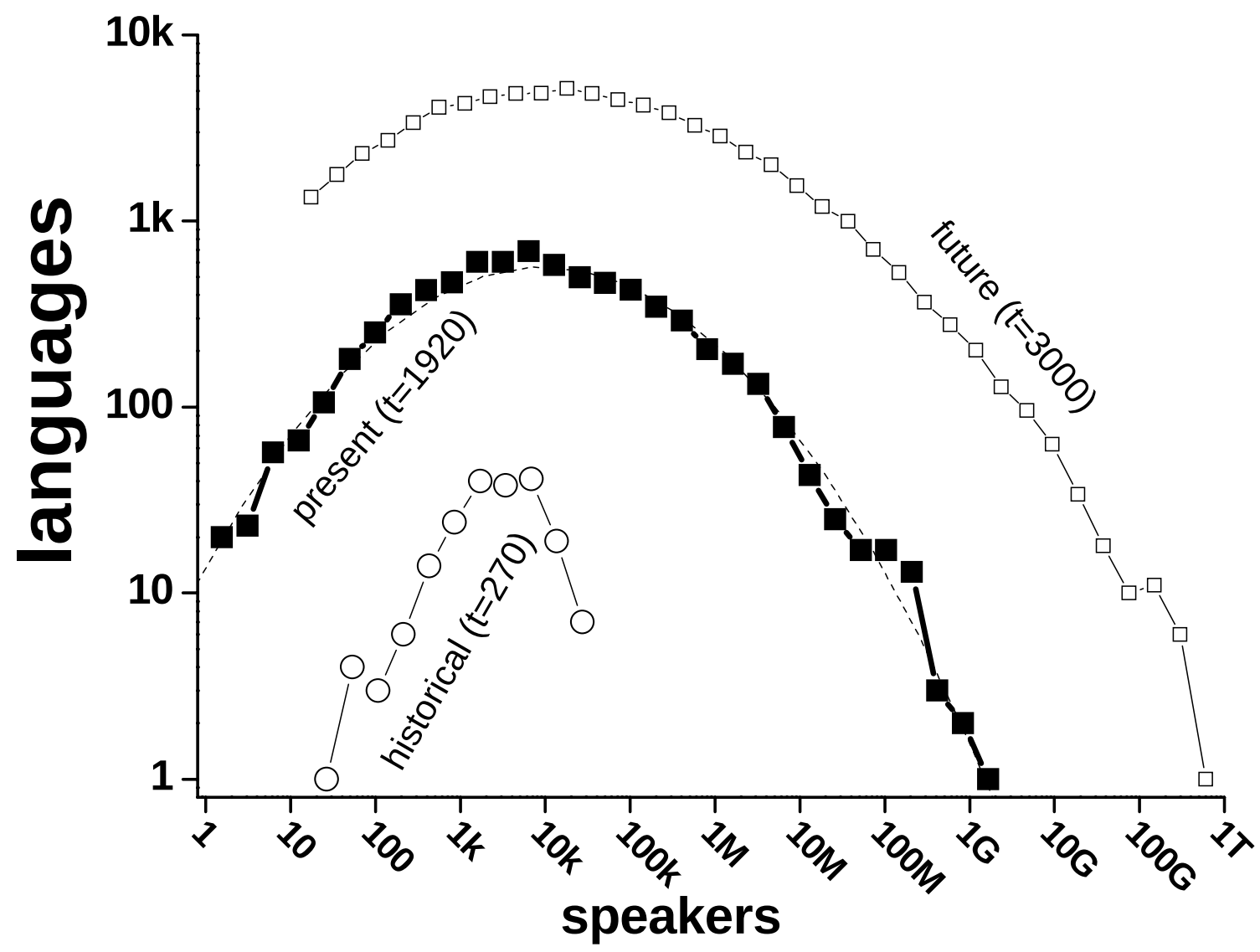

**Figure 2** Historical (open circles, for $t$=270) and present (squares, for $t$=1920) size distribution of living languages. (Parameters are given in sect. 3.1.) Dashed line is a parabolic fit indicating that, the present distribution is slightly asymmetric log-normal. For some other theoretical plots and for the empirical data; see, [1-3].) Outer plot is for future ($t$=1920).

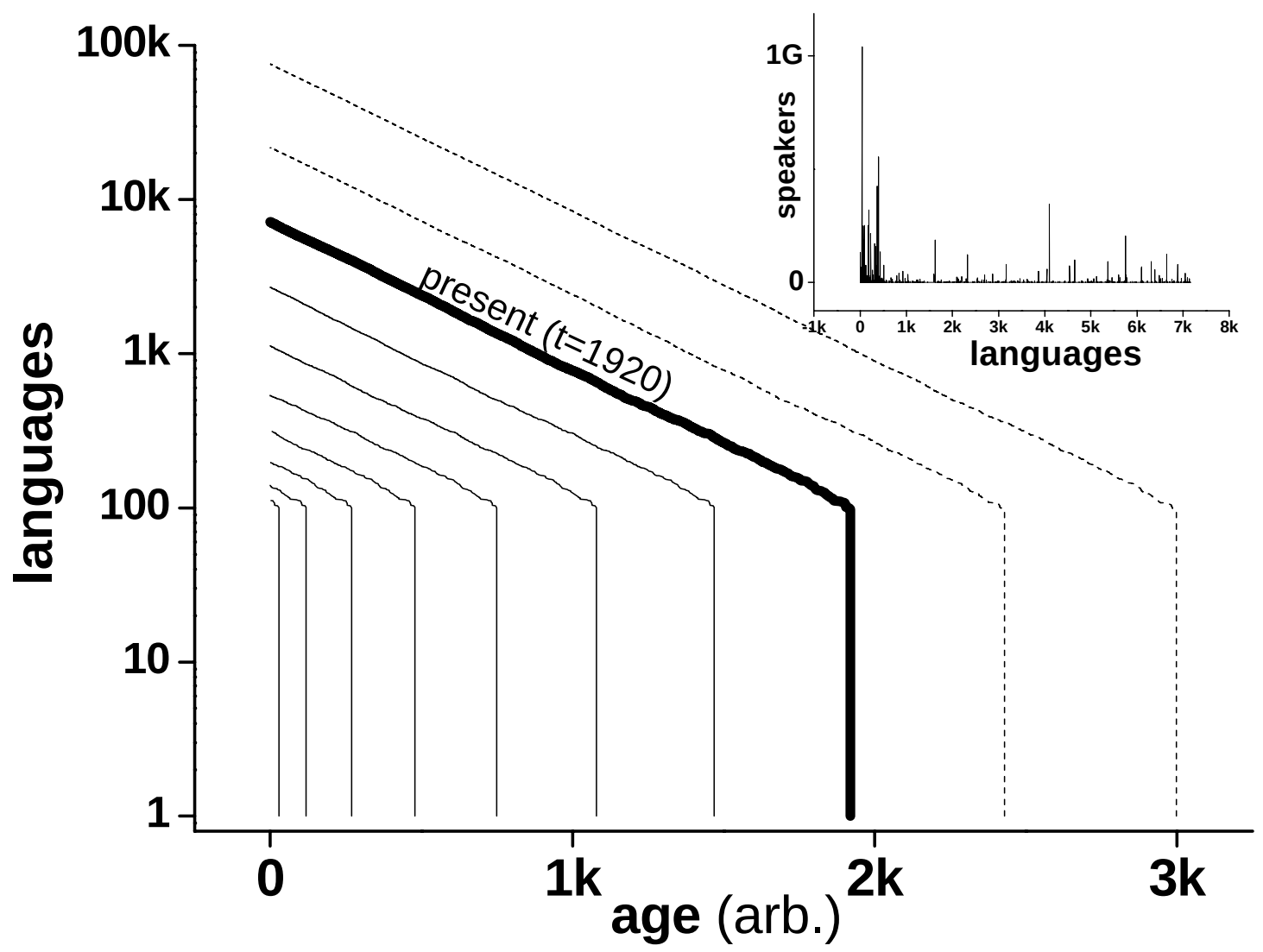

**Figure 3** Ages of living languages at present ($t$=1920, thick solid), past ($t$<1920, thin solid) and in future (1920<$t$, dashed). (Parameters are given in sect. 3.1.) Please note that, the number of generated (new) languages decreases exponentially with age, in all cases. Inset is the number of speakers for each of the living language at present time (which are evolved out of the initial ones), where the biggest language (Chinese) is spoken by little more than one billion (1G, in the inset) human. Please note also that, many of the ancestor languages are still living, and few (only two) of them (out of initial ones, $M(0)$=100) became extinct in terms of fragmentation.

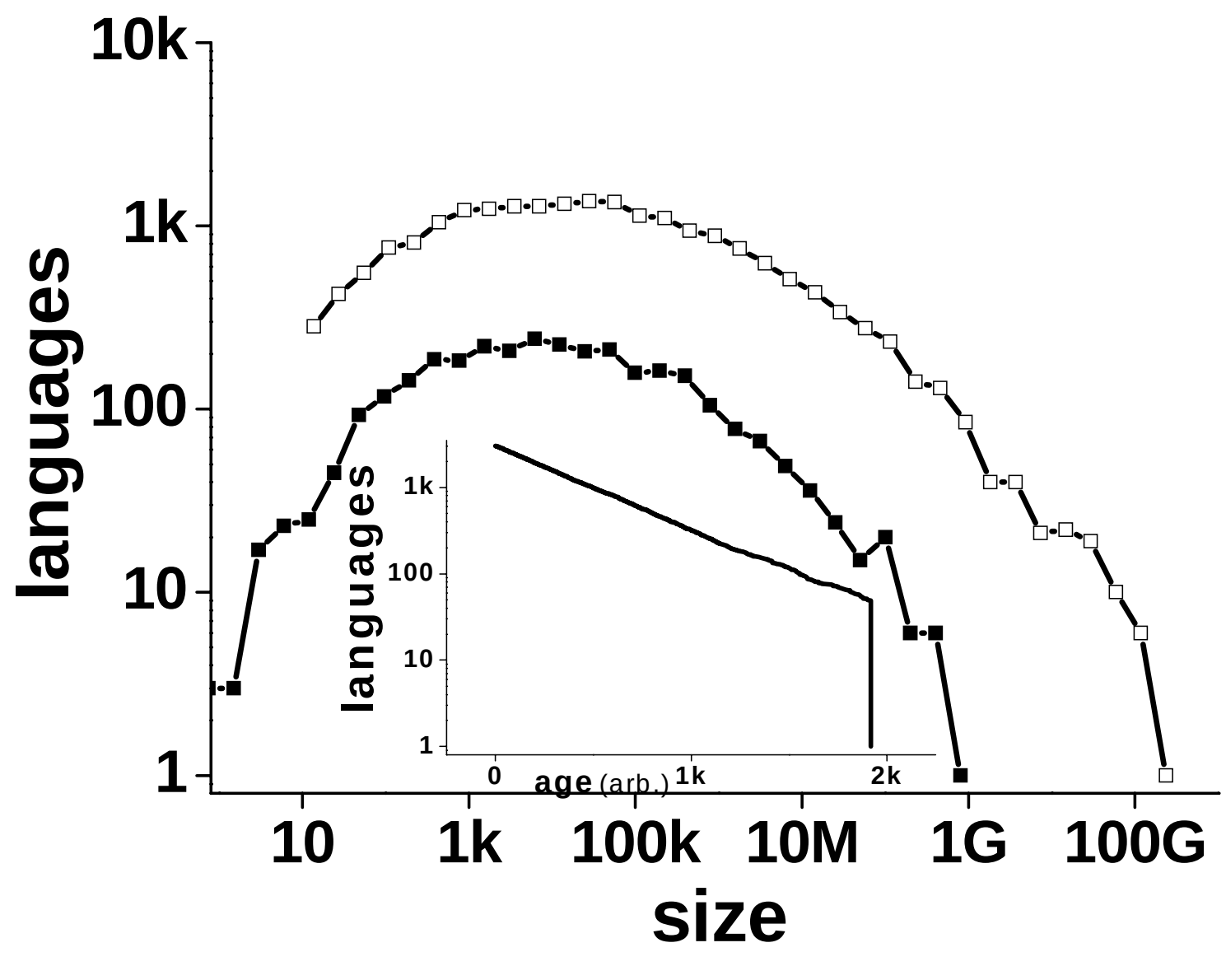

**Figure 4** Size distributions of languages: Solid squares are for the present time ($t$=1920) with $G$ ($\neq$1), where evolution takes place under light elimination of languages (See, the relevant text for the other parameters.); and open squares are for future ($t$=2430) with the same elimination $G$ ($\neq$1) (other parameters are same as in Fig. 3.) Inset is the age distribution with $G$, at $t$=1920, where one may observe that, many ancestor languages are eliminated, and slope (exponent for the age distribution) is smaller than the ones given in Fig. 3 for $G$ =1.

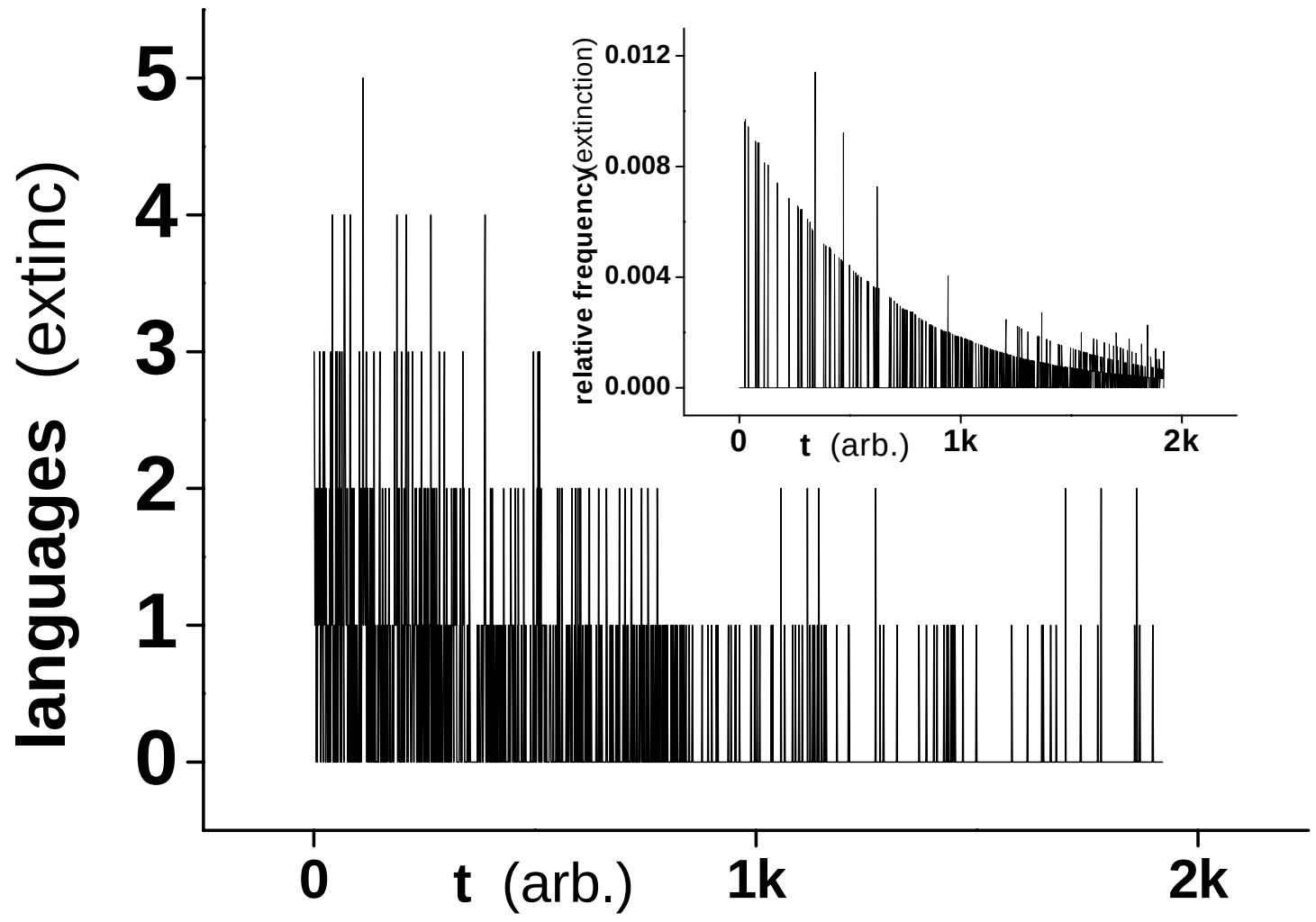

**Figure 5** Time distribution for lifetime of the languages, which became extinct; and, the inset is relative frequency of events for extinction, all with *G*<1 (where other parameters are same as before).

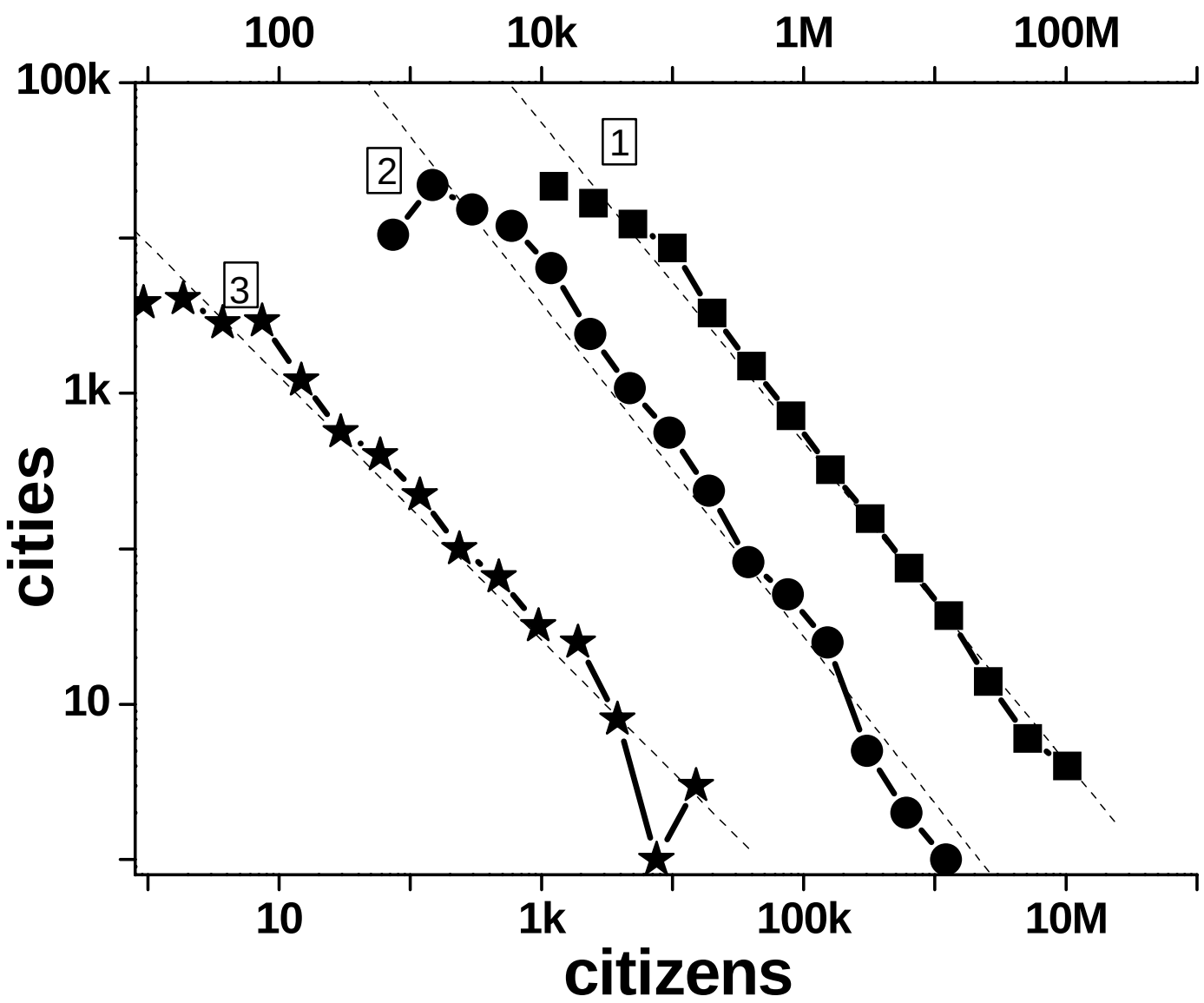

**Figure 6** Power laws for cities. Slopes of the linear fits are: 1) (squares) -1.03 (with left and bottom axes); 2) (circles) -1.07 (with left and top axes); 3) (stars) -0.85 (with left and bottom axes) Relevant parameters are given in sect. 3.2.

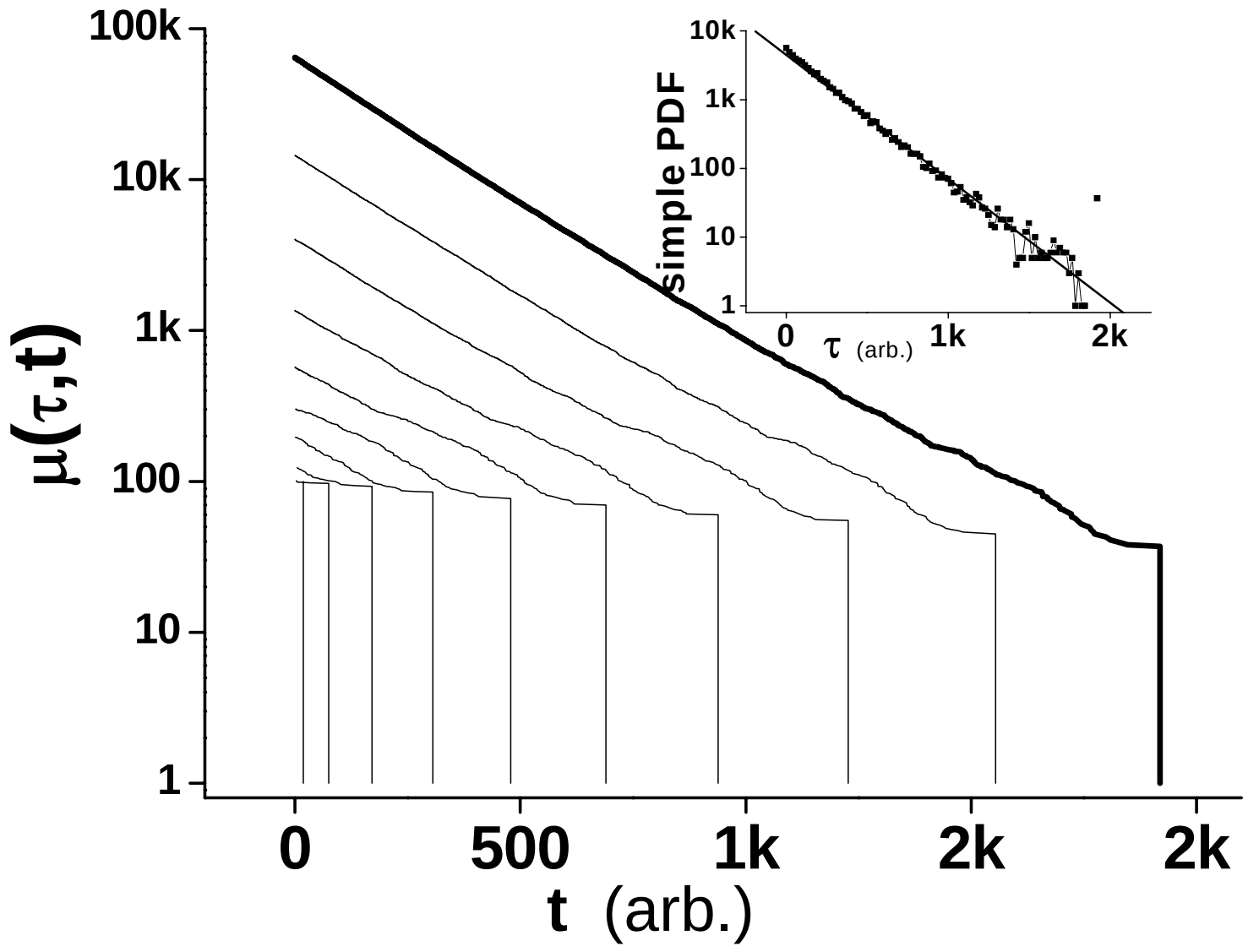

**Figure 7** Time distribution for (extinct and living) cities. Inset is the simple probability distribution function for t=1920, where the slope of linear fit is -0.00185. (Parameters are: $S$=0.1, $H$= 0.015, $G$=0.9996, and minimum population of a city is 514; others are same as Fig. 5.)

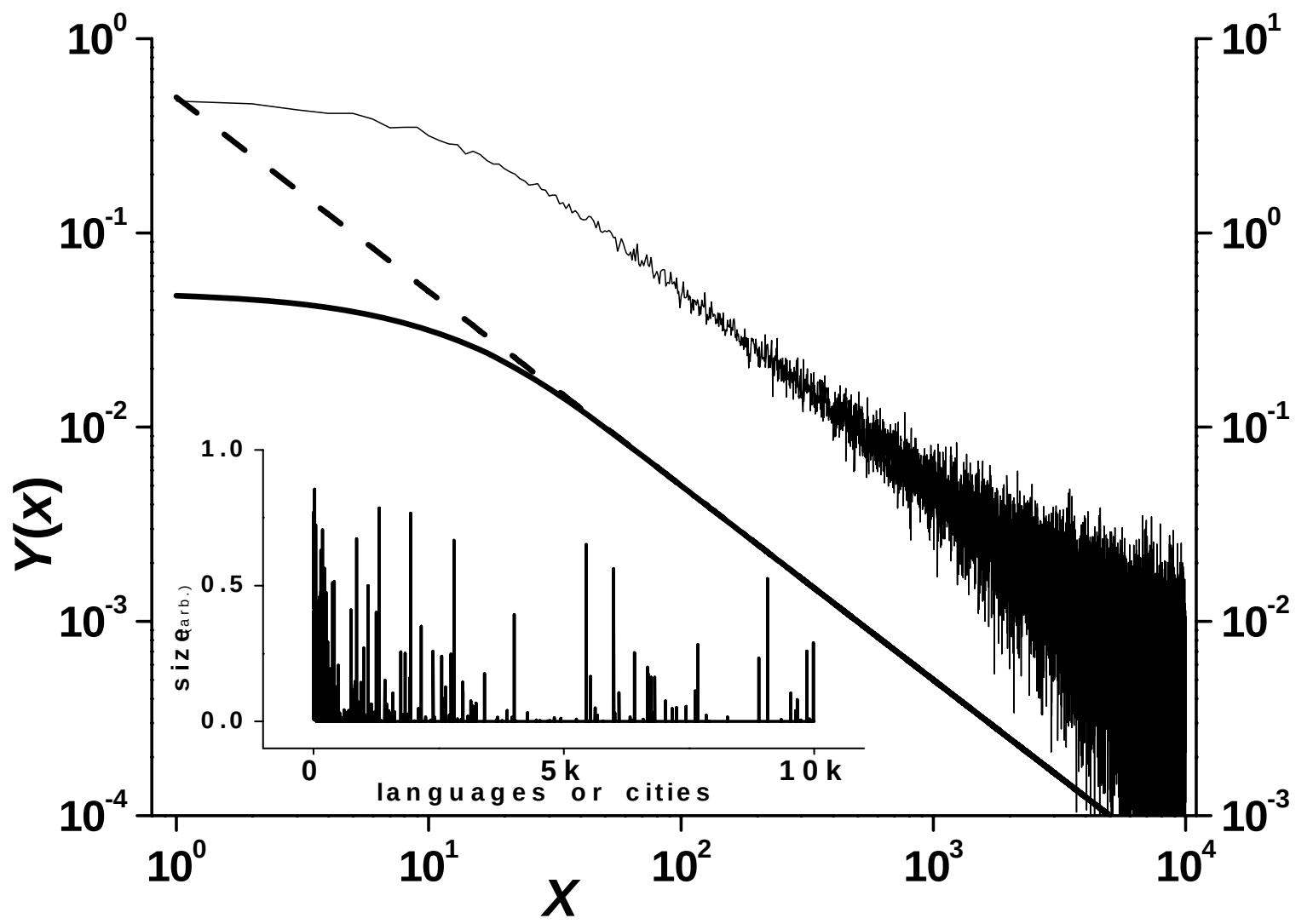

**Figure 8** $Y(x)$ (Eqs. (4-8)) with $I=J=1000$, for $A_{min}= B_{min} =0.0$, $A_{max} \leq 1.0$, $B_{max} \leq 1.0$, where the oscillating plot is for Eq. (4) with the vertical axis on left, and with the vertical axis on right are: Solid plot, for the analytical expression in Eq. (7); and the dashed line, for Eq. (8). Please note that, vertical axes are shifted with respect to each other, yet units are same. Secondly, solid and dashed line has slope minus unity for large $x$. It is obvious that, oscillating plot displays power law minus unity (Pareto-Zipf law) for large $x$. Inset is $y_{ij}(x)$ in Eq. (3), where two independent random numbers are used; one for exponent, and one for amplitude. It is clear that $y_{ij}(x)$ may be utilized for size of languages or cities along $x$-axis, where many ancestors (for small $x$) are present. (See, the inset in Fig. 4, which is for 10 thousand agents.)